\documentclass[aps,pre,twocolumn,superscriptaddress]{revtex4-1}
\usepackage{amsmath}
\usepackage{amssymb}
\usepackage{graphicx}
\usepackage{color}
\begin{document}
\title{Influence of Bonded Interactions on Structural Phases of  
Flexible Polymers}
\author{Kai Qi}
\email{k.qi@fz-juelich.de}
\affiliation{Theoretical Soft Matter and Biophysics, Institute of Complex
Systems and Institute for Advanced Simulation, Forschungszentrum J\"ulich,
D-52425 J\"ulich, Germany}
\affiliation{Soft Matter Systems Research Group, Center for Simulational
Physics, The University of Georgia, Athens, GA 30602, USA}
\author{Benjamin Liewehr}
\email{Benjamin.Liewehr@uni-rostock.de}
\affiliation{Soft Matter Systems Research Group, Center for Simulational
Physics, The University of Georgia, Athens, GA 30602, USA}
\affiliation{Institute of Physics, University of Rostock, 
Albert-Einstein-Stra{\ss}e 23, D-18059 Rostock, Germany}
\author{Tomas Koci}
\affiliation{Soft Matter Systems Research Group, Center for Simulational
Physics, The University of Georgia, Athens, GA 30602, USA}
\author{Busara~Pattanasiri}
\email{faasbrp@ku.ac.th}
\affiliation{Soft Matter Systems Research Group, Center for Simulational
Physics, The University of Georgia, Athens, GA 30602, USA}
\affiliation{Department of Physics, Faculty of Liberal Arts and Science,
Kasetsart University, Kamphaeng Saen Campus, Nakhon Pathom
73140, Thailand}
\author{Matthew J. Williams}
\email{mwilliams72@murraystate.edu}
\affiliation{Soft Matter Systems Research Group, Center for Simulational
Physics, The University of Georgia, Athens, GA 30602, USA}
\affiliation{Institute of Engineering, Murray State University, Murray, 
KY 42071, USA}
\author{Michael Bachmann}
\email{bachmann@smsyslab.org; http://www.smsyslab.org}
\affiliation{Soft Matter Systems Research Group, Center for Simulational
Physics, The University of Georgia, Athens, GA 30602, USA}
\date{\today}
\begin{abstract}
We introduce a novel coarse-grained bead-spring model for flexible polymers 
to systematically examine the effects of an adjusted 
bonded potential on the formation and stability of structural macrostates in a
thermal environment.
The density of states obtained in advanced 
replica-exchange Monte 
Carlo simulations is analyzed by employing the recently developed 
generalized microcanonical 
inflection-point analysis method, which enables the identification of diverse 
structural phases and the construction of a suitably parameterized
hyperphase diagram. It reveals that icosahedral phases dominate for 
polymers with
asymmetric and narrow bond potentials, whereas 
polymers with symmetric and more elastic bonds tend to form amorphous 
structures 
with non-icosahedral cores. We also observe a hierarchy 
in the freezing transition behavior associated with 
the formation of the surface layer after nucleation.
\end{abstract}
\maketitle 
%
%
%
\section{Introduction}
Biological functions and processes of biopolymers such as DNA and proteins are 
inevitably connected to their geometric structures. Numerous studies in 
interdisciplinary research have been initiated by the necessity for a better 
understanding of dynamical and structural properties of polymers. Given the
complexity of the chemical structure of biomolecules, only experimental and 
computational studies can help to gain deeper insights into the physical 
mechanisms guiding cooperative, qualitative changes of the
system's macrostate. Essential is the understanding of the response of the 
polymer system to modifications of external and intrinsic parameters and the  
formation of stable, and potentially functional, structural 
phases~\cite{flory1,deGennes1,grosberg1,doi1,dillBook1,Michael2014Book}.

Generic, coarse-grained polymer models prove to be extremely helpful for 
the thermodynamic analysis of structural phases of polymers. In simulations 
of effective models on mesoscopic scales, model parameter spaces can be 
scanned efficiently, which is hardly possible in 
specific, microscopic models, whose parametrization typically requires 
hundreds of force-field parameters. Therefore, generic models offer 
the option of a systematic and better understanding of the general 
structural behavior for entire classes of polymers.

One of the general questions in this context is how interaction ranges and 
symmetries affect the structure formation processes and the 
thermodynamic stability of conformational phases of polymers. It could be 
shown that the stability and existence of a globular or liquid phase in 
models of flexible polymers depends on the
effective interaction range between nonbonded 
monomers~\cite{Taylor20091,Taylor20092,Gross2013,Werlich2017}. A 
recent study of the 
influence of bond confinement upon
structural phases and transition behavior of a flexible chain showed that the 
liquid phase also disappears with increasing bond fluctuation range and the 
gas-liquid and the liquid-solid transition lines merge~\cite{Koci2015}. 
Bending restraints in helical polymers can lead to the formation of stable 
helix bundles, resembling tertiary structures in 
biopolymers~\cite{Williams2015}.

However, the influence of the energy scale of the potential 
between bonded monomers 
on thermodynamic and geometric features of structural phases of the 
polymer has not yet been 
thoroughly addressed. In this paper, we study the effects of bonded  
interactions 
on the conformational behavior of a flexible elastic homopolymer. 
The
interaction strength between bonded monomers is adjusted by 
a parameter 
that controls the shape of the bonded potential. 
For the sampling of the conformational state space, parallel 
tempering 
simulations~\cite{SwendsenWangPT,GeyerPT,HukushimaNemotoPT,HansmannPT} were 
performed, 
supported by a 
parallel version of multicanonical 
sampling~\cite{Michael2014Book,Berg1991,Berg1992,Janke1998,Berg2000,%
Berg2003,Zierenberg2013}. 

We use the recently introduced generalized microcanonical inflection-point 
analysis method~\cite{qb1}, which has been designed to 
systematically identify and classify 
phase transitions.
Based on the 
results obtained in the simulations,
the hyperphase diagram parameterized by 
the temperature and the bond potential parameter is 
constructed. For the understanding of the freezing transition process and 
the discussion of structure types dominating the solid phases, a systematic 
structural analysis is performed as well.

The paper is organized as follows. Our novel versatile coarse-grained model 
for flexible polymers,
the simulation methods, and the statistical 
analysis techniques employed are described 
in Sec.~\ref{model_simulation_methods}.  
In Sec.~\ref{results}, we discuss the results of the canonical, 
microcanonical, and structural analyses. The 
paper is concluded by the summary in Sec.~\ref{summary}.
%
%
\section{Polymer Model, Simulation, and Statistical 
Analysis Methods}
\label{model_simulation_methods}
%
\subsection{Versatile Model for Flexible Polymers}
\label{model}
In this study, we systematically investigate the influence of the 
shape and effective range of the potential between bonded monomers 
in a generic model of a self-interacting flexible elastic 
homopolymer with 55 monomers. This ``magic'' chain length has been chosen 
because of the stable icosahedral ground-state conformation in conventional 
models for flexible polymers~\cite{svbj1,sbj1}. We assume the 
molecular  
interaction between non-bonded monomers is of van der Waals type and 
can be modeled 
by a standard 12-6 Lennard-Jones (LJ) potential,
\begin{equation}
 U_\textrm{LJ}(r) = 4\epsilon\left[\left(\frac{\sigma}{r}\right)^{12} - 
\left(\frac{\sigma}{r}\right)^{6}\right],
\end{equation}
where $\sigma$ is the van der Waals radius and $r$ denotes the 
monomer-monomer distance. 
For computational efficiency, the non-bonded potential 
is truncated at $r_\textrm{c} = 2.5 \sigma$ and shifted by the 
constant $U_\text{shift} = U_\textrm{LJ}(r_\textrm{c})$ to avoid a 
discontinuity of the potential at $r=r_\textrm{c}$,
\begin{equation}   
U_\textrm{NB}(r) = \left\{
\begin{array}{ll} 
U_\textrm{LJ}(r) - U_\textrm{shift}, & r < r_\textrm{c}, \\
0, & \text{otherwise}.
\end{array}\right.
\label{U_nb}
\end{equation}
The minimum of the potential is located at $r_0=2^{1/6} \sigma$, 
which fixes the length scale associated with this interaction. 
In the simulations we set $r_0 \equiv 1$.

The elastic bond between two adjacent monomers is modeled by the combination 
of the
finitely extensible nonlinear elastic (FENE) 
potential~\cite{fene1,fene2,fene3} and a Lennard-Jones potential,
\begin{eqnarray}
U_\textrm{B}(r) &=& -\frac{1}{2} \kappa R^2 \ln\left[1-
\left(\frac{r - r_0}{R} \right)^2\right]  \nonumber \\ 
&&+ \eta\, [U_\textrm{LJ}(r) + \epsilon] - (U_\textrm{shift}+\epsilon),
\label{U_b}
\end{eqnarray}
where the bond potential parameter $\eta$ controls the width and 
asymmetry of the potential.

\begin{figure}
\includegraphics[width =\columnwidth]{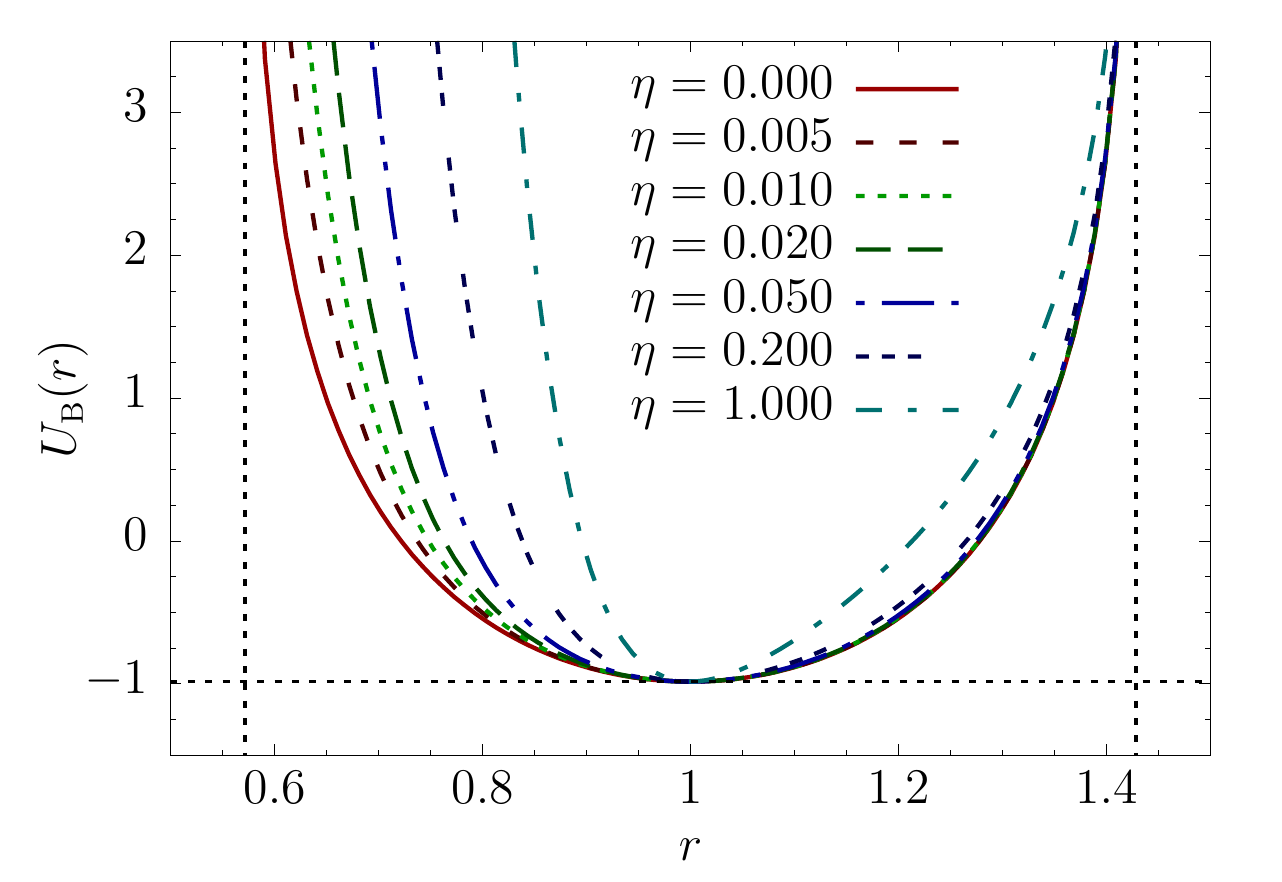}
\caption{\label{Fig_U_B} Potential between bonded monomers $U_\mathrm{B}(r)$ 
for different values 
of the parameter $\eta\in[0,1]$, which we
introduced to control the range and shape of the bonded 
potential.}
\end{figure}

The maximum bond extension is limited by the FENE potential, which 
diverges for $r \rightarrow r_0 \pm R$.
The FENE parameters are set to standard values $R \equiv 3/7$ and 
$\kappa \equiv 98/5$~\cite{Qi2014}. The Lennard-Jones potential of the bonded 
interaction 
causes an asymmetry for $\eta > 0$ and reduces the potential width, 
while the location of the minimum $r_0=1$ is unchanged for this 
choice of parameters. The potential is shifted to have 
the same minimum value as in the non-bonded case:
$U_\textrm{B}(r_0) = -\epsilon - U_\textrm{shift} = U_\textrm{NB}(r_0)$. 
Figure~\ref{Fig_U_B} shows the bonded potentials for various 
values of $\eta$.
We chose $\epsilon$ as the overall energy scale and set it to unity in the 
simulations. 

In our model, the total energy of a conformation for a polymer with $N$ 
monomers, $\textbf{X} = (\mathbf{r}_1,\ldots,\mathbf{r}_{N})$,
is thus given by
\begin{equation}
E(\textbf{X}) = \sum\limits_{j=1}^{N-2} \sum\limits_{k=j+2}^{N}  
U_\textrm{NB}(r_{jk}) + \sum\limits_{j=1}^{N-1} U_\textrm{B}(r_{j\, j+1}),
\label{E_sum}
\end{equation}
where $r_{jk}=|\mathbf{r}_j-\mathbf{r}_{k}|$ is the distance between 
monomers $j$ and $k$.
%
\subsection{Simulation Method}
\label{simulation_method}
We performed parallel tempering 
simulations~\cite{SwendsenWangPT,GeyerPT,HukushimaNemotoPT,HansmannPT}
for $21$ different values of $\eta\in[0,1]$ to measure thermodynamic 
and structural quantities needed for the structural analysis.
In our implementation of this replica-exchange Monte Carlo method, 
Metropolis sampling was performed at 
temperatures $T_i\in [0.11, 3.00]$ with $i=1,2,\ldots,I$. The number of 
temperature threads $I$ ranged from 96 to 128, depending 
on the level of complexity imposed on the system by the different 
choices of the model parameter $\eta$. At each temperature, 
random displacement 
updates $\mathbf{r} \rightarrow \mathbf{r} + \mathbf{d}$ were performed 
within a cubic box with edge lengths $l_i > |d_n|, n =1,2,3$. 
The size of the box $l_i$ was adjusted for every temperature 
thread separately prior to the simulation in order to achieve a 
Metropolis acceptance rate of approximately $50\%$.
To facilitate the decorrelation of structures, exchanges of 
replicas between neighboring temperatures were proposed and 
accepted with the exchange probability 
\begin{equation}
\hspace*{-2.5mm}P(E_i,\beta_i;E_{i+1},\beta_{i+1}) = 
\min\left(1, e^{(E_i-E_{i+1})(\beta_i - \beta_{i+1})}\right),
\end{equation}
where $\beta=1/k_\mathrm{B} T$ (with $k_\mathrm{B}\equiv 1$ in the 
simulations).
The temperatures $T_i$ were chosen equally distant 
in $\beta$ space to guarantee sufficiently high exchange 
probabilities in the low-temperature regime, where large autocorrelation 
times are expected~\cite{Qi2014}. Between each of the $10^6$ replica 
exchanges, $10^3$ Metropolis Monte Carlo sweeps were performed. 

For each ensemble generated in the parallel-tempering simulation,
an estimate of the density of states $\overline{g}_i(E)$ was 
calculated from the energy histogram of the single-temperature 
canonical ensemble $h_i(E)$, utilizing 
$\overline{g}_i(E)=h_i(E) e^{\beta E}$. The estimated density of states 
calculated for a particular value of $\beta_i$ is a relative quantity 
which is only reliable in the energy region that is 
adequately covered by the ensemble of states statistically representative at 
this temperature. We used 
the multi-histogram reweighting method~\cite{Swendsen1989,Rosenberg1992} to 
combine the estimates of the
density of states obtained from the different temperature threads. 
The estimator $\hat{g}(E)$ for the density of states covering the 
whole energy space sampled is thus given by
\begin{equation}
\hat{g}(E)=\frac{\sum_{i=1}^{I} h_i(E)}{\sum_{i=1}^{I} 
M_i \hat{Z}_i^{-1} e^{-\beta_i E}}, 
\end{equation}
where $M_i$ is the total number of sweeps and
$\hat{Z}_i$ the individual partition function estimator in the $i$-th thread,
\begin{equation}
\hat{Z}_i = \sum_E \hat{g}(E) e^{-\beta_i E}.
\end{equation}
This system of equations is iterated until convergence is achieved.

We also performed parallel simulations of multicanonical 
sampling~\cite{Berg1991,Berg1992,Janke1998,Berg2000,Berg2003,Zierenberg2013,%
Michael2014Book} as a supportive method for the structural analysis of the 
solid phase. 
Multicanonical sampling is more effective at overcoming hidden free-energy 
barriers associated with first-order transitions than parallel tempering 
schemes which do not artificially enhance the sampling of entropically 
suppressed regions in phase space.
This combination of simulation methods 
enabled the verification of the simulation
results achieved by parallel tempering and the estimates for the 
density of states to be used in the microcanonical statistical analysis. 
%
\subsection{Generalized Microcanonical Inflection-Point Analysis Method}
\label{micro_analysis}
For the identification and classification of the structural polymer phases, 
we use the recently developed generalized microcanonical inflection-point 
method~\cite{qb1}. This statistical analysis method consequently combines 
microcanonical thermodynamics~\cite{Gross2001} and the principle of minimal 
sensitivity~\cite{Stevenson1981,Stevenson1981b}, and generalizes earlier 
approaches~\cite{Schnabel2011} limited to low-order transitions. As it turns 
out, though, higher-order phase transitions are potentially relevant as well, 
but have widely been ignored in conventional canonical analyses.
Typical first- and second-order indicator functions (order parameters and 
response functions) are not sensitive enough to expose higher-order 
transition 
signals.

The general idea is to focus on the essential quantities entropy and energy, 
which govern any possible macroscopic behavior of a physical system 
in response to environmental conditions. In other words, the response is 
already encoded in the system's energetic and entropic properties. 
Consequently, entropy can be defined microcanonically as the logarithm of 
the available energetic phase space. This makes entropy dependent of energy 
and the sole function that governs the system behavior. It is common to 
introduce the microcanonical entropy, up to an irrelevant constant, by the 
logarithm of the density of states,
\begin{equation}
S(E)=k_\mathrm{B}\ln\, g(E),
\label{eq:microent}
\end{equation}
where $k_\mathrm{B}$ is the Boltzmann constant. Alternatively, the 
integrated density of states $\int_{E_0}^E dE'\,g(E')$ (integrating from the 
ground-state energy $E_0$) can be used as well as a representative of the 
energetic phase-space volume, but both versions lead to virtually identical 
results near the most interesting features, phase transitions, as entropy 
varies rapidly in these energy regions. We use the simpler 
form~(\ref{eq:microent}) in the following.

The microcanonical entropy and its derivatives have a well-defined monotonic 
behavior within energy regions associated with a single phase. A transition 
between significantly different macrostates, however, disturbs the monotony 
and is signaled by inflection points. Referring to the 
principle of minimal sensitivity, only least-sensitive inflection points have 
a physical meaning, though. Including the derivatives of $S(E)$ in this 
consideration 
enables the introduction of a systematic identification and classification 
scheme for phase transitions~\cite{qb1}, reminiscent of Ehrenfest's idea of 
using thermodynamic potentials~\cite{ehrenfest1}. However, microcanonical 
inflection-point analysis is not based on the 
response of the system to changes of external thermodynamic state variables 
such as the (canonical) temperature.

If $S(E)$ contains a least-sensitive inflection point, it must clearly be 
signaled by a minimum in the first derivative, which is the microcanonical 
temperature
\begin{equation}
\beta(E)=T^{-1}(E)=d S(E)/d E.
\end{equation}
We classify this as a first-order transition signal. In consequent 
analogy, we define second-order transitions by means of least-sensitive 
inflection points in 
$\beta(E)$. If a least-sensitive inflection point occurs 
first in the second derivative, which is symbolized by 
$\gamma(E)=d^2 S(E)/d E^2$, we refer to it as a third-order transition, etc. 

Generally, transitions of odd order $(2\nu-1)$ ($\nu$ positive integer) 
possess a least-sensitive inflection point in the $(2\nu-2)$th derivative of 
$S(E)$, which is characterized by a positive-valued minimum in 
\begin{equation}
\left. \frac{d^{(2\nu-1)}S(E)}{dE^{(2\nu-1)}} 
\right|_{E=E_{\mathrm{tr}}}>0. 
\label{odd_order_trans}
\end{equation}
Transitions of even order $2\nu$ are least sensitive at the transition energy 
in the $(2\nu-1)$th derivative of $S(E)$ and the corresponding maximum in the 
$(2\nu)$th  derivative must be negative:
\begin{equation}
\left. \frac{d^{2\nu}S(E)}{dE^{2\nu}} 
\right|_{E=E_{\mathrm{tr}}}<0. 
\label{even_order_trans}
\end{equation}
We call this class of transitions \emph{independent}, 
because there is another category, 
\emph{dependent transitions}, which can only occur as 
precursors of independent transitions~\cite{qb1}. In this paper, we can focus 
on independent transitions as dependent transitions were not observed for 
this model.

Note that due to binning, i.e., the discretization of the energy space in the 
simulations, the numerical results for the entropy are discrete. Employing 
discrete differentiation methods naively to get the derivatives needed 
for the subsequent microcanonical analysis would enhance the noise 
associated with the numerical error of the data and obscure transition 
signals. Therefore, we use the discrete set of 
$Q+1$ data points $S_q$ at energies $E_q$ (with $q=0,1,\ldots,Q$) obtained in 
the simulations as control points for the B\'ezier 
algorithm~\cite{Bezier1968,Gordon1974,Michael2014Book}, which generates the 
smooth function
\begin{equation}
S_\mathrm{bez}(E)= \sum_{q=0}^{Q} \binom{Q}{q} \!
\left(\frac{E_Q-E}{E_Q-E_0}\right)^{Q-q}\!
\left(\frac{E-E_0}{E_Q-E_0}\right)^{q} S_q.
\label{bezier_smoothing}
\end{equation}
The derivatives required for the statistical analysis can be 
calculated from this function.

Although we do not expect to observe divergences 
in the transition region for finite systems, we do find clear signals 
indicating 
structural transitions in the microcanonical curves, which serve as 
finite-system analogs of 
classical phase transitions. The order of these structural transitions 
is discernible from subtle details of the relationship 
between microcanonical entropy or its derivatives and the energy
of the system. This approach is not based on catastrophic behavior of 
thermodynamic quantities. Consequently, it does not require a scaling 
analysis in the artificial
thermodynamic limit of infinitely large systems, which for many realistic 
systems (such as biopolymers) is not even possible. 
%
%
\section{Results}
\label{results}
%
\subsection{Canonical Statistical Analysis}
\label{can_analysis}
Canonical analysis is the conventional approach to understanding 
thermodynamic properties of a system. Extremal thermal fluctuations of 
any observable $O$, defined by 
\begin{equation}
\frac{d}{d T}\langle O\rangle = \frac{\langle O E 
\rangle-\langle O\rangle\langle E\rangle}{k_{\rm B}T^2},
\label{eq:fluctuation}
\end{equation}
are used to locate regions of temperature space with enhanced 
thermal activity. The most commonly considered observable is 
the system energy $E$, which is readily available from the simulations since 
Monte Carlo simulations require its calculation after each update. The
thermal fluctuation of the energy, the heat capacity 
$C_V(T)=d\langle E\rangle/dT$, is a
useful generic indicator for transitions in complex systems. Peaks in the  
$C_V(T)$ curve signal rapid changes of $\langle E\rangle(T)$, which typically 
accompany significant macrostate changes of the system like in thermodynamic 
phase 
transitions.
More specific for locating structural transitions in polymer systems, 
we also estimated the squared radius of gyration,
\begin{equation}
R^2_{\mathrm{gyr}}  = \frac{1}{N}\sum_{j=1}^{N} \left(
\mathbf{r}_j - \mathbf{r}_\mathrm{c.m.}\right)^2,
\label{eq:Rg2}
\end{equation}
where $\mathbf{r}_j$ is the coordinate of monomer $j$ and 
$\mathbf{r}_\mathrm{c.m.}= \sum_{j=1}^{N}\mathbf{r}_j/N$ is the center of mass 
of the polymer conformation. The radius of 
gyration can be interpreted as a measure for structural compactness and, 
therefore, helps to distinguish structural phases in which the macrostates 
notably differ in size. 
The corresponding thermal fluctuation quantity
$d\langle R_{\mathrm{gyr}}^2 \rangle / dT$ 
is particularly useful for the identification of structural
transitions if $C_V$ fails to provide a pronounced signal. This typically 
occurs for entropy-driven transitions in small systems that do not exhibit 
large energetic fluctuations. The most prominent example is the coil-globule 
(or $\Theta$) transition in finite polymer systems.

\begin{figure}
\centerline{\includegraphics[width =\columnwidth]{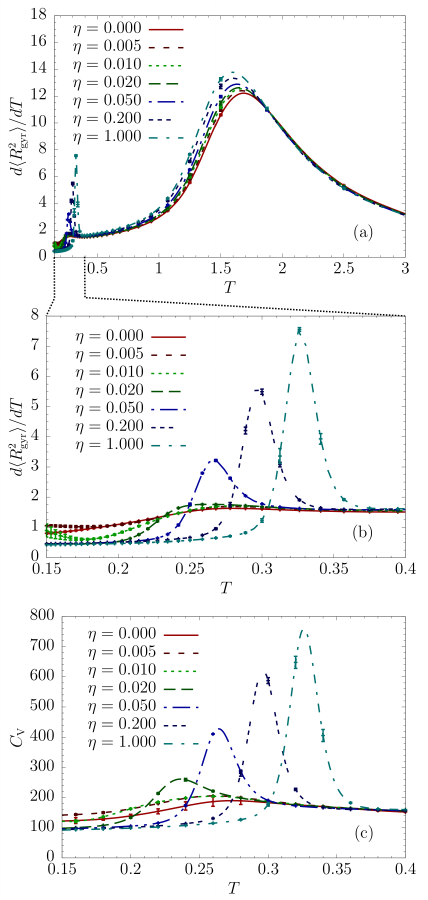}}
\caption{\label{canonical_quantities} (a) and (b) 
Thermal fluctuation of the squared 
radius of gyration $d\langle R_{\mathrm{gyr}}^2\rangle / dT$ as a function 
of temperature for a flexible 
polymer with $N = 55$ monomers at different 
values of $\eta$, (c) low-temperature region of the heat capacity $C_V$.}
\end{figure}

Figure~\ref{canonical_quantities} shows the plots of $d\langle 
R_{\mathrm{gyr}}^2 \rangle / dT$  and $C_V$
of the 55-mer as functions of temperature $T$ for various $\eta$ 
values, respectively. The clearly visible peaks in the fluctuations of 
structural 
compactness at $T\approx 1.6$ in Fig.~\ref{canonical_quantities}(a) 
indicate the $\Theta$ transition, where extended coil structures collapse 
and form the more compact globules. Increasing $\eta$ slightly 
shifts the $\Theta$ transition point to lower temperatures. 

By lowering the temperature further, globular structures eventually freeze 
into solid conformations. The 
corresponding transition signals can be observed in the group of peaks 
at temperatures around $T=0.3$ [Figs.~\ref{canonical_quantities}(b, c)]. 
These 
transition signals
shift to lower temperatures for small $\eta$ values but start to move up 
in temperature for $\eta \geq 0.2$.
This observation suggests a significant change in system behavior if $\eta$ 
exceeds a certain threshold value.
Among these visible features, the 
signals with narrow widths and high peak heights at $\eta \geq 0.2$ 
indicate the freezing transition. These signals become less pronounced 
and broader as $\eta$ decreases. Instead of being 
indicators for a specific type of transition, these wide and low 
peaks are rather envelopes of multiple transition signals. This ambiguity 
in distinguishing and classifying the transitions at small $\eta$ values 
is caused by finite-size effects which cannot be resolved by 
means of canonical statistical analysis of $C_V$ and $d\langle 
R_{\mathrm{gyr}}^2 \rangle / dT$. Therefore, it is necessary to employ other 
systematic and robust methods such as microcanonical 
inflection-point 
analysis, which can clearly distinguish the sensitive 
transition signals in finite-size systems. Subsequent structural analysis 
will then enable us to interpret the 
physical meaning of these transition signals. 
%
%
\subsection{Microcanonical Analysis and\\ Hyperphase Diagram}
\label{micro_results}
\begin{figure*}
\centerline{\includegraphics[width =\textwidth]{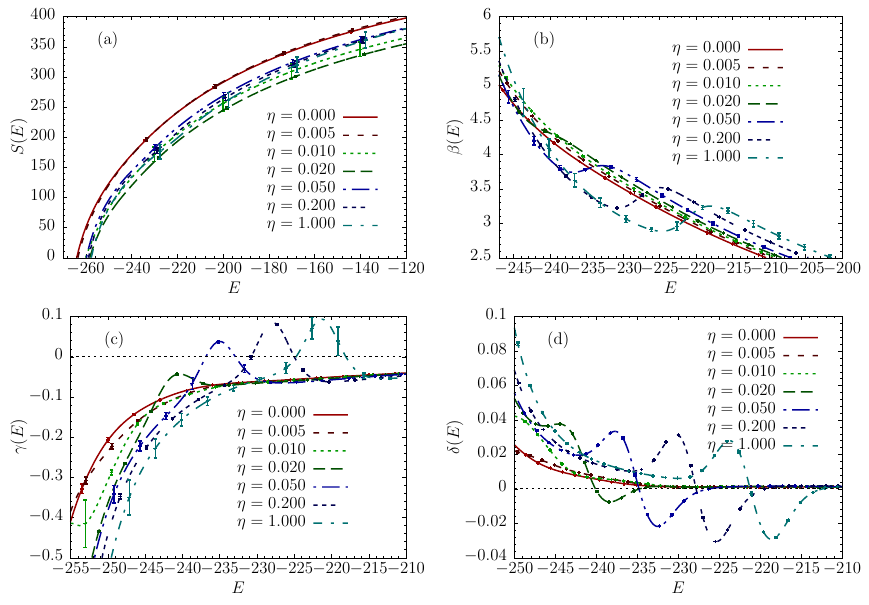}}
\caption{\label{microcanonical_quantities} (a) Microcanonical
entropy $S(E)$ for an array of $\eta$ values; (b) microcanonical 
inverse temperature $\beta(E)=dS/dE$; (c) $\gamma(E)=d\beta(E)/dE$; (d) 
$\delta(E)=d\gamma(E)/dE$.}
\end{figure*}
\begin{figure}
\centerline{\includegraphics[width =\columnwidth]{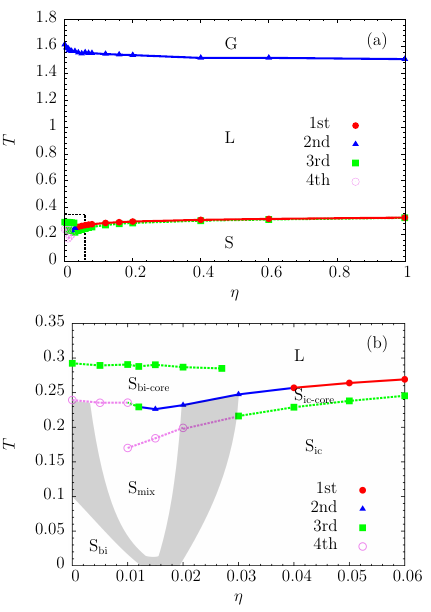}}
\caption{\label{phase_diagram}%
(a) Microcanonical hyper-phase 
diagram, parameterized by temperature $T$ and model 
parameter $\eta$. The major phases are labeled G (gas), L (liquid), and S 
(solid). (b) Detail view of the low-temperature 
and small-$\eta$ region, emphasizing the different solid phases. 
In S$_{\mathrm{bi-core}}$, structures with bihexagonal cores and 
liquid-like shells are dominant. S$_{\mathrm{ic-core}}$ represents 
conformations with well-formed icosahedral cores, but incomplete surface 
layer. In the S$_{\mathrm{ic}}$ and S$_{\mathrm{bi}}$ pseudophases, 
icosahedral and bihexagonal core structures have complete overlayers. In 
S$_{\mathrm{mix}}$ 
icosahedral and bihexagonal core structures coexist. The 
``solid'' subphases are separated by gray empirical transition bands.
Dashed lines represent lines of transitions higher than second order.}
\end{figure}

As discussed earlier, microcanonical inflection-point 
analysis~\cite{Schnabel2011,Michael2014Book,qb1} can be used to identify and 
classify transitions in systems of any size. In the following, we focus on 
the 
transition behavior of polymer models with
$\eta\in[0,1]$ at low temperatures. Plots of the microcanonical entropy 
$S(E)$ are shown in Fig.~\ref{microcanonical_quantities}(a). Simulations for 
$\eta\ge 0.04$ reveal least-sensitive inflection points and a convex region 
in the entropy curves, leading to the prominent ``backbending behavior'' in 
the $\beta(E)$ plots shown in Fig.~\ref{microcanonical_quantities}(b). 
According to the generalized inflection-point analysis method~\cite{qb1}, 
minima of $\beta(E)$ are indicative of a first-order transition. At 
$\eta= 0.04$, the first-order transition is found at $T\approx0.25$. It is 
stable for $\eta \ge 0.04$ and the transition temperature 
increases slightly with $\eta$. This is the expected first-order transition 
from the liquid to a solid phase, which is known to be the
icosahedral phase for the 55-mer studied here. These results enable us to 
construct the low-temperature transition line in the 
hyperphase diagram shown in Fig.~\ref{phase_diagram}(a).

Below $\eta\approx 0.04$, the scenario is significantly different. The 
negative-valued peak for $\eta=0.02$ at $E\approx-241$ corresponds to a 
least-sensitive inflection point in $\beta(E)$, which indicates a 
second-order transition at $T\approx 0.23$. Systematic analysis reveals that 
this transition type occurs in the interval $0.01<\eta<0.04$ and replaces the 
first-order liquid-solid transition in this temperature region. As the 
structural analysis in the following section will show, the liquid-solid 
transition behavior becomes indeed more complex. This is also due to the 
occurrence of additional transition signals of higher order. At $\eta\approx 
0.012$ we find a third-order transition signal in this transition region, 
which marks the crossover point towards a fourth-order transition for 
$\eta<0.01$. 

Another remarkable feature of the small-$\eta$ models is the occurrence of 
additional precursor lines that ``shadow'' the major low-$T$ transition 
line. Their existence is unmistakably manifest from the positive-valued 
minima in the $\delta(E)$ curves shown in 
Fig.~\ref{microcanonical_quantities}(d), which indicate third-order 
transitions. For $\eta>0.025$, these transition points lie below the 
liquid-solid transition line, for $\eta<0.025$ above the extension of this 
line. This qualitative change is another indication that the structural 
behavior of the system significantly changes around this crossover point. It 
is worth noting that the shadow transition keeps accompanying the liquid-solid 
line even for larger $\eta$ values. It approaches the strong first-order line 
asymptotically and merges with it (in microcanonical analysis, a transition 
line is swallowed by a first-order transition, if the former enters the 
backbending region of the latter).

Based on the results obtained by microcanonical inflection-point analysis for 
19 $\eta$ 
values, we construct the hyperphase diagram as shown
in Fig.~\ref{phase_diagram}(a). The major phases are well-known from 
previous studies of flexible polymers. At sufficiently high temperatures, the 
polymer 
is in the gas-like pseudophase in which dissolved or random-coil 
structures dominate. Decreasing the temperature causes the polymer to 
collapse and to enter the liquid pseudophase, where  
compact globular conformations are favorably formed. The 
corresponding pseudophase transition is the well-known 
$\Theta$-transition (collapse transition). Because of the 
negative $\gamma(E)$ peaks for all $\eta$ values in this region, this 
transition 
is classified as of second order and it is represented by the blue line in 
Fig.~\ref{phase_diagram}(a). As the temperature decreases further, 
the polymer transfers from the globular phase to the more compact 
``solid'' phase (liquid-solid or freezing transition), which is 
characterized by locally crystalline or 
amorphous metastable structures. 

The analysis of the compact subphases is much more challenging as  
the magnification of the low-temperature 
and small-$\eta$ region of the hyper-phase diagram in 
Fig.~\ref{phase_diagram}(b) shows. Multiple transition lines of 
different order exist. 
Microcanonical 
inflection-point analysis identifies the liquid-solid transition as of first 
order for $\eta>0.04$ and of 
second order for $0.01<\eta<0.04$. The extension of this line for
$\eta<0.01$ is of higher than second order. The detailed structural analysis 
in the next section will shed more light on the reasons for this diverse and 
complex behavior, which includes additional transition lines. Most notably, 
the liquid-solid transition is accompanied by a third-order transition line. 
Both lines eventually merge for large $\eta$ values. Structures in the 
intermediate phase possess a solid icosahedral core, whereas the overlayers 
are still disorganized and incomplete, forming a two-dimensional liquid 
interface. Only for temperatures below this companion line and sufficiently 
large $\eta$ values, the phase is dominated by structures with icosahedral 
geometry as it is expected for a flexible polymer with a ``magic'' number of 
monomers and matching length scales of bonded and nonbonded interactions. 

For $\eta<0.03$, different geometric subphases exist and are characterized by 
the competition between structures with icosahedral and bihexagonal cores 
(see Fig.~\ref{fig:core_structures}). Only for smallest $\eta$ values and 
lowest temperatures, the bihexagonal phase dominates. This is an interesting 
result as it shows that the solid phase of polymers with symmetric and 
wider bond potentials is indeed different, but stable only for small 
variations of parameters. If the threshold value $\eta\approx 0.02$ is 
exceeded, though, icosahedral structures mix in and eventually dominate for 
larger parameter values.
%
\subsection{Structural analysis}
\label{stucture_analysis}
The tools of microcanonical inflection-point analysis, as 
introduced in the previous section, provide us with a systematic 
way of identifying and classifying all structural transitions in 
a given physical system. Another step towards a more advanced 
understanding of thermodynamic properties of a system is the 
identification of dominant conformations and their abundance 
in a relevant energy range. This can be done directly 
either by visual inspection of sample structures, or more 
systematically, by introducing a suitable set of structural order 
parameters.
\begin{figure}
\centerline{\includegraphics[width = 0.8\columnwidth]{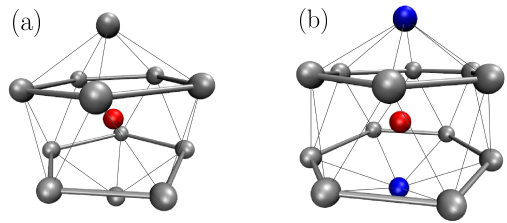}}
\caption{Core structures in solid phases of the 55-mer: (a) icosahedral 
(ic-core); (b) bihexagonal (bi-core).} 
\label{fig:core_structures} 
\end{figure}
\begin{figure*}
\centerline{\includegraphics[width = 0.9\textwidth]{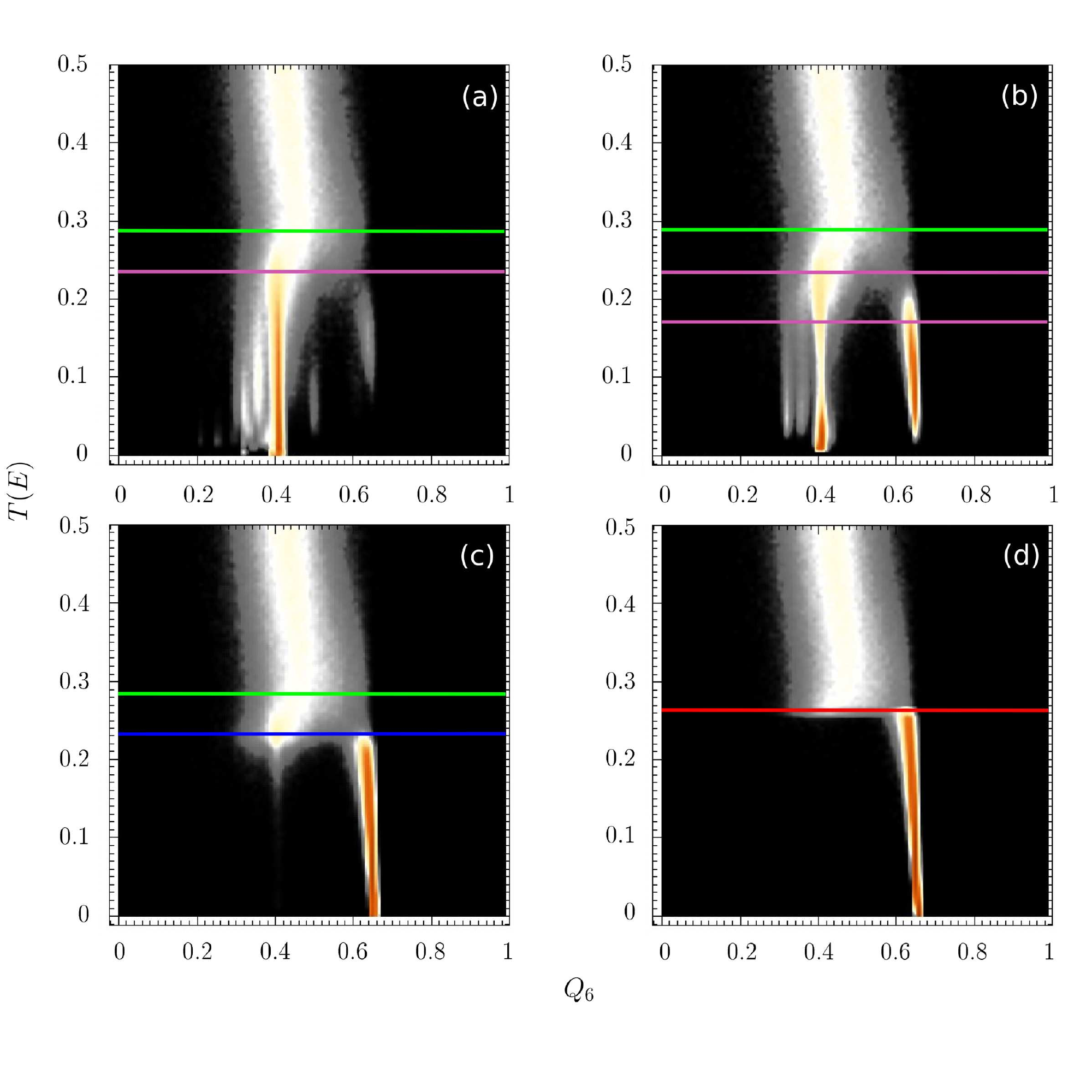}}
\caption{Intensity plots of the $Q_6$ order parameter 
at (a) $\eta = 0.005$, (b) $\eta = 0.010$, (c) $\eta = 0.020$, 
and (d) $\eta = 0.050$, respectively. The probability 
of finding polymer structures with a particular value of $Q_6$ is 
represented by colored shading, with black being zero probability and red 
corresponding to the maximum probability of $1$.}
\label{fig:intensity_plots}
\end{figure*}

For the purpose of identification of low-energy solid-like structures 
which possess well-defined symmetries, a set of effective order parameters 
can be defined in terms of the real spherical 
harmonics~\cite{kqb1}. We define the polymer core as 
consisting of $K$ monomers with 
the coordinates 
$\mathbf{X}^\mathrm{C} = 
(\mathbf{r}^\mathrm{C}_1,\ldots,\mathbf{r}^\mathrm{C}_K)$
within 1.25$\sigma$ of the 
center of mass of the polymer.
We introduce a set of rotationally invariant order parameters by 
\begin{equation}
Q_{l} = \left[\frac{4\pi}{2l + 1} \sum_{m = -l}^{l}
|\rho_{lm}|^{2} \right]^{1/2},
\end{equation}
where 
\begin{equation}
\rho_{lm} = \frac{1}{K} \sum\limits_{k=1}^{K}
Y_{lm}(\mathbf{r}^\mathrm{C}_k)
\label{eq:15}
\end{equation}
is the average of the real spherical harmonics
\begin{equation}
\hspace*{-2.5mm}
Y_{lm}(\mathbf{r}) = 
\begin{cases}
\vspace{1mm}
\frac{i}{\sqrt{2}}\left[Y_{l}^{m}(\mathbf{r}) - (-1)^{m} 
Y_{l}^{-m}(\mathbf{r}) \right], & m < 0,\\[1mm]
Y_{l}^{m}(\mathbf{r}), &  m = 0, \\[1mm]
\frac{1}{\sqrt{2}}\left[Y_{l}^{-m}(\mathbf{r}) + (-1)^{m} 
Y_{l}^{m}(\mathbf{r})\right], &  m > 0
\end{cases}
\end{equation}
calculated at the positions 
of the core monomers. 

As expected, preliminary inspection of structures obtained from 
simulations of the flexible 55-mer with $\eta \approx 1$ 
shows that below the freezing transition virtually all conformations
contain an icosahedral core. However, with virtually symmetric bonded 
interactions $(\eta < 0.03)$, two distinct core 
geometries are found. In addition to the standard icosahedral core, 
which is present in the global minimum structures of most short chains, 
we have also identified a bihexagonal core consisting of 15 monomers 
(Fig.~\ref{fig:core_structures}). The six-fold dihedral symmetry of 
the bihexagon and the icosahedral symmetry are best distinguished using 
the $Q_6$ order parameter. For a perfect icosahedral core, 
$Q_6 \approx 0.65$, whereas a bihexagonal core corresponds to 
$Q_6 \approx 0.41$.  

\begin{figure}
\centerline{\includegraphics[width = \columnwidth]{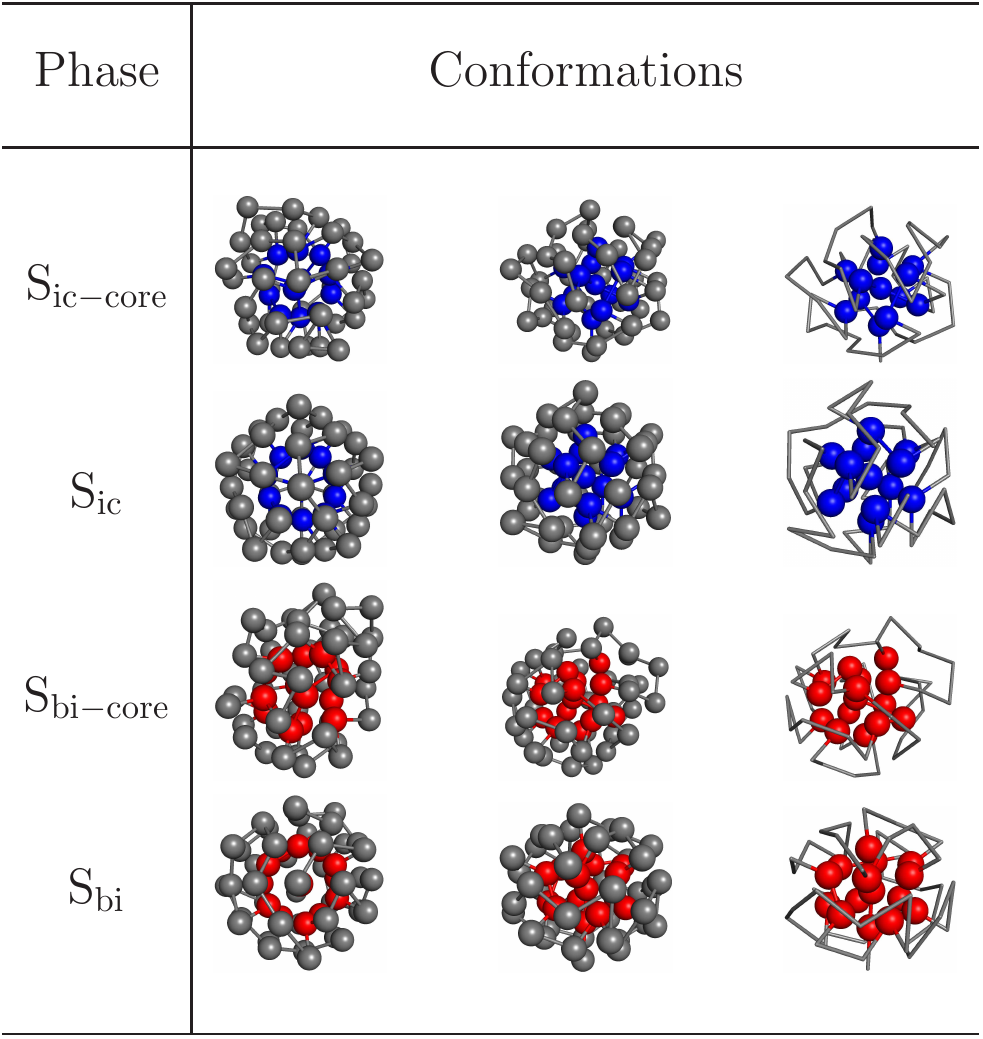}}
\caption{Different views of typical structures in the solid subphases: view 
along the core axis (left), perspective view (center), and core 
representation (right). Icosahedral and bihexagonal cores are plotted in blue 
and red, respectively. Gray beads represent surface monomers.}
\label{fig:configurations}
\end{figure}

We present the results in the form of intensity plots in 
Fig.~\ref{fig:intensity_plots}. Shades correspond to the probability of finding a 
structure with a particular value of $Q_6$ at a given microcanonical 
temperature $T(E)$. Black represents zero probability and red unity.
An interesting feature, found only in systems with $\eta < 0.027$, is marked by the 
green horizontal lines at $T\approx 0.29$. It is associated with the apparent 
shift of the peak of the $Q_6$ distribution at this temperature towards 
lower values. This indicates the onset of the formation of bihexagonal 
cores in the liquid phase and corresponds to the third-order transition 
line in the microcanonical phase diagram [Fig.~\ref{phase_diagram}(b)]. Below this 
transition, amorphous structures with loose bihexagonal cores and 
liquid-like surfaces are identified in the S$_{\mathrm{bi-core}}$ 
pseudophase. Typical conformations containing icosahedral or bi-hexagonal 
cores are shown in Fig.~\ref{fig:configurations}.

At low temperatures and $\eta = 0.005$, we observe a single dominant funnel 
centered at $Q_6\approx 0.41$, containing structures with a bihexagonal 
core [Fig.~\ref{fig:intensity_plots}(a)]. 
The adjacent secondary funnels all contain bihexagonal cores with 
slightly modified inter-monomer distances. Structures with an 
icosahedral core are found in the weakly populated funnel at 
$Q_6\approx 0.65$. The low-temperature transition signal is 
associated with the increase in the population of structures with 
bihexagonal cores and is classified as of fourth order. This transition 
signal is marked by violet transition lines 
in Fig.~\ref{fig:intensity_plots}(a) and in the hyper-phase diagram 
shown in Fig.~\ref{phase_diagram}(b). A small increase in the strength 
of the bonded LJ potential leads to a sharp increase in the population 
of icosahedral cores. For $\eta = 0.01$, the ground state of the 
polymer is still found in the bihexagonal funnel, however the onset 
of a significant population of icosahedral cores produces an additional 
fourth-order transition signal at $T = 0.17$, see Fig.~\ref{phase_diagram}(b) 
and Fig.~\ref{fig:intensity_plots}(b). Further increase leads to a sharp 
decline in the population of the bihexagonal funnel. 

\begin{figure}
\centerline{\includegraphics[width = \columnwidth]{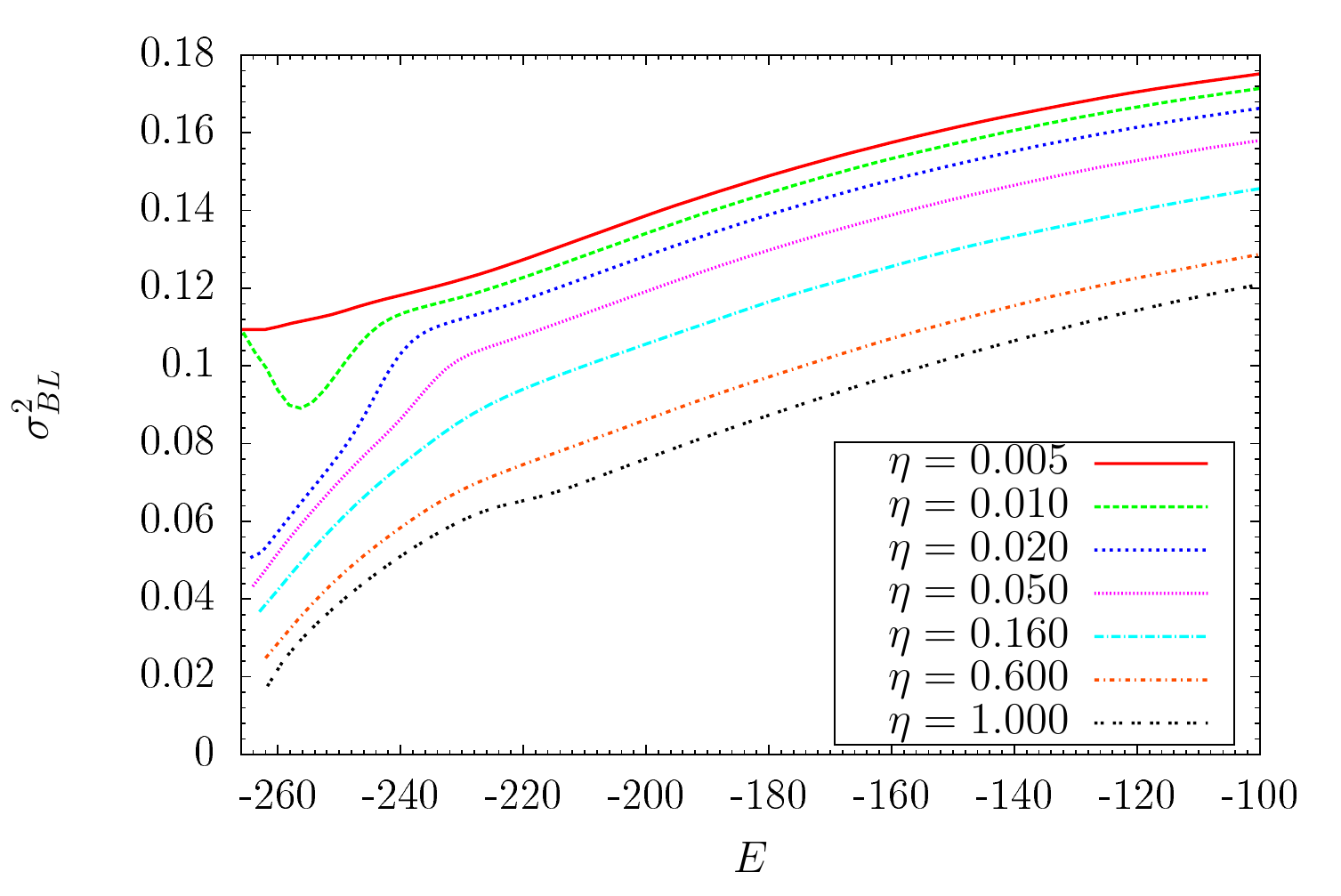}}
\caption{Variance $\sigma^2_\mathrm{BL}$ of bond-length fluctuations for the 
55-mer plotted as a 
function of energy and different values of the model parameter $\eta$. }
\label{fig:bond_length_variance} 
\end{figure}

This can be seen clearly for $\eta \geq 0.02$ 
[Fig.~\ref{fig:intensity_plots}(c), (d)] where the energetic penalty 
for non-optimal bond lengths becomes too large to accommodate structures 
with bihexagonal cores. Indeed, their formation requires significant 
variance in bond lengths, whereas icosahedral cores can be formed with 
near-optimal values. For $\eta\approx 0$, the pure FENE potential permits 
large bond fluctuations. However, with the introduction of the bonded LJ 
potential these fluctuations cause an energetic penalty. This explains 
why the bihexagonal funnel exists only if the bonded LJ potential is 
sufficiently weak. In Fig.~\ref{fig:bond_length_variance}, we show the 
bond-length variance as a function of energy for different values of the model 
parameter $\eta$. With increasing values of $\eta$ the variance decreases, most 
significantly in the low-energy region. Most striking is the difference 
between the low-energy curves for $\eta = 0.01$ and $\eta = 0.02$, where 
the former has a bihexagonal ground-state and the latter is icosahedral. 

At $\eta \approx 0.04$, the signal associated with the onset of 
the icosahedral funnel becomes first order and can be unambiguously 
identified as the freezing transition. Beyond $\eta\approx 0.1$, the 
structural and energetic properties of the 55-mer do not change significantly 
anymore.

From the structural analysis, we conclude that
the low-temperature behavior of this flexible polymer model can
be classified into three pseudophases. For sufficiently small $\eta$ 
values, bihexagonal core structures with compact
hexagonal-like surface layers, shown in 
Fig.~\ref{fig:configurations}, dominate the solid phase. As $\eta$ is 
increased, the population of 
icosahedral structures gradually increases
due to the additional energy penalty of the non-optimal bond 
lengths in bihexagonal cores, but both structure types still coexist in a 
mixed phase [$S_{\mathrm{mix}}$ in the
Fig.~\ref{phase_diagram}(b)]. Once the width of the bonded potential 
is sufficiently narrow, the bihexagonal core structures completely 
vanish and only the icosahedral structures persist. 
The corresponding pseudophase is labeled $S_{\mathrm{ic}}$ in 
Fig.~\ref{phase_diagram} and representative
structures are shown in Fig.~\ref{fig:configurations}.

The third-order transition accompanying the freezing transition in 
Fig.~\ref{phase_diagram}(b) is associated with the
completion of the icosahedral shell. Upon decreasing the temperature, 
liquid structures begin to nucleate and first form the stable 
icosahedral cores if $\eta>0.03$. Surfaces of the polymer structures 
still undergo large fluctuations in order to arrange the surface 
monomers in optimal locations. The mobility of the monomers is confined 
to effectively two dimensions on the surface. These semi-liquid structures 
are dominant in the solid $S_{\mathrm{ic-core}}$ pseudophase. 
If the temperature is decreased further, the surface formation 
finishes and complete icosahedral shell structures appear in the 
solid $S_{\mathrm{ic}}$ pseudophase. 

For sufficiently large $\eta$ values, the freezing 
transition of a polymer can generally be characterized by two hierarchical 
processes. One is associated 
with the nucleation process of the core identified as first- or second-order 
transitions and the other is due to surface layer formation which is a 
third-order transition.
%
%
\section{Concluding Remarks}
\label{summary}
We employed parallel tempering simulations, supported by a parallelized 
version of multicanonical sampling, to investigate the effects 
of the shape of the potential of bonded monomers on the structure formation 
properties of elastic flexible polymers.
For this purpose, we introduced the model 
parameter $\eta$, which controls the width and asymmetry of the
bond potential. In this study we focused on a single flexible polymer with 
55 monomers, which is an interesting example as it can form a perfectly 
icosahedral structure under certain conditions. In order to identify and 
distinguish the 
various structural phases in this system, we systematically applied the 
microcanonical inflection-point analysis method and performed a thorough 
structural analysis of the compact phases to 
construct the hyperphase diagram.  

Besides the commonly expected random-coil (gas-like) and globular 
(liquid-like) phases, we find a diversity of solid subphases whose 
properties depend on the value of the $\eta$ model parameter. 

Perturbing the symmetric FENE potential allows for larger fluctuations of
bond lengths. Structures with bihexagonal cores
commonly dominate the solid phase as long as the bond potential is 
virtually symmetric. Increasing the value of the model parameter $\eta$ 
narrows the bond potential width and induces asymmetry.  
The energetic penalty for non-optimal bond lengths becomes too large 
to accommodate structures with bihexagonal cores. Thus, bihexagonal 
core structures become less favorable for large $\eta$ values and 
icosahedral cores begin to dominate, leading to a mixed phase, in which both 
structure types coexist. The mixed phase 
eventually turns into the icosahedral phase if $\eta$ is sufficiently large 
and the bond potential is 
dominated by the Lennard-Jones potential. 

Our results also indicate that for sufficiently large $\eta$ values, the 
freezing transition is a well-organized hierarchical process: Whereas 
nucleation or core formation is a 
first-order transition, the subsequent shell formation process 
was identified as a separate third-order transition. During core 
formation, the surface layer remains liquid. This flexibility enables an 
optimal arrangement of the core monomers. Once the solid core is formed, the 
monomers of the surface layer pack optimally in the void spaces left on the 
surface of the core, thereby forming the second shell of the 
icosahedral conformation. 

The results we obtained also demonstrate the power 
of microcanonical inflection-point analysis, which does not only help
identify the major transitions but can also distinguish the details of 
the transition processes by signaling higher-order transitions.

Our case study of a 55-mer provides robust insights into the nature 
of transition processes in flexible polymers. The general 
structure of the hyperphase diagram discussed in this paper is not expected 
to change significantly for 
larger systems. However, it is well known that details of the liquid-solid and 
solid-solid 
transitions, typically associated with Mackay and anti-Mackay overlayer 
formation~\cite{sbj1},
depend on the system size. Therefore, future work on the deeper analysis of 
these processes for other chain lengths, using the model and methodologies 
introduced in this paper, would be intriguing.
%
%
\begin{acknowledgments}
This study has been supported partially by the NSF under Grant No.\
DMR-1463241. B.P.\ was supported by The Royal Thai
Government Scholarship under the Development and Promotion of Science and 
Technology Talent Project (DPST). B.L.\ acknowledges support by the German 
Academic Scholarship Foundation.
\end{acknowledgments}
%
%

%
\end{document}